\documentclass[sigconf]{acmart}

\AtBeginDocument{%
  }
\newtheorem{definition}{Definition}

\usepackage{ifthen}
\usepackage[ruled,vlined]{algorithm2e}
\usepackage{cleveref}
\usepackage{subcaption}

\newcommand{\IfThenElse}[3]{%
  \ifthenelse{#1}{%
    \begin{quote}#2\end{quote}%
  }{%
    \begin{quote}#3\end{quote}%
  }%
}
\acmConference[2024]{31th ACM Conference on Computer and Communications Security (CCS)}{October 14--18,
  2024}{Salt Lake City, U.S.A.}

\copyrightyear{2024}
\acmYear{2024}
\setcopyright{rightsretained}
\acmConference[CCS '24]{Proceedings of the 2024 ACM SIGSAC Conference on Computer and Communications Security}{October 14--18, 2024}{Salt Lake City, UT, USA}
\acmBooktitle{Proceedings of the 2024 ACM SIGSAC Conference on Computer and Communications Security (CCS '24), October 14--18, 2024, Salt Lake City, UT, USA}
\acmDOI{10.1145/3658644.3691411}
\acmISBN{979-8-4007-0636-3/24/10}

\begin{document}

\title{Poster: Protection against Source Inference Attacks in Federated Learning using Unary Encoding and Shuffling}


\author{Andreas Athanasiou$^*$}
\email{andreas.athanasiou@inria.fr}
\orcid{0009-0000-3251-2148}
\affiliation{
  \institution{INRIA and LIX, IPP}
  \city{Palaiseau}
  \country{France}
}

\author{Kangsoo Jung$^*$}
\email{gangsoo.zeong@inria.fr}
\orcid{0000-0003-2070-1050}
\affiliation{
  \institution{INRIA and LIX, IPP}
  \city{Palaiseau}
  \country{France}
}

\author{Catuscia Palamidessi}
\email{catuscia@lix.polytechnique.fr}
\orcid{0000-0003-4597-7002}
\affiliation{%
  \institution{INRIA and LIX, IPP}
  \city{Palaiseau}
  \country{France}
}

\renewcommand{\shortauthors}{Andreas Athanasiou, Kangsoo Jung, \& Catuscia Palamidessi}
\begin{abstract}
Federated Learning (FL) enables clients to train a joint model without disclosing their local data. Instead, they share their local model updates with a central server that moderates the process and creates a joint model.  However, FL is susceptible to a series of privacy attacks.  Recently, the source inference attack (SIA) has been proposed where an honest-but-curious central server tries to identify exactly which client owns a specific data record. 

In this work, we propose a defense against SIAs by using a trusted shuffler, without compromising the accuracy of the joint model. We employ a combination of unary encoding with shuffling, which can effectively blend all clients' model updates,   preventing the central server from inferring information about each client's model update separately. In order to address the increased communication cost of unary encoding we employ quantization. 
Our preliminary experiments show promising results; the proposed mechanism notably decreases the accuracy of SIAs without compromising the accuracy of the joint model.

\end{abstract}





\maketitle
\def\thefootnote{*}\footnotetext{Primary authors with equal contribution.
\\
© {A. Athanasiou | ACM} {2024}. This is the author's version of the work. It is posted here for your personal use. Not for redistribution. The definitive Version of Record was published in Proceedings of the 2024 ACM SIGSAC Conference on Computer and Communications Security, https://doi.org/10.1145/3658644.3691411

}\def\thefootnote{\arabic{footnote}}

% \Af{Andreas: You can add coments like this} 
% \Kf{Kangsoo: Like this} 
% \Cf {Catuscia: Like this}

\section{Introduction}

In FL \cite{fl_mother_paper}, each client independently trains a model using their own data and then sends the model update to a central server. The server aggregates these model updates to create a new joint model, which is then distributed back to the clients. The process continues iteratively for multiple rounds, usually until the model converges. However, in a naive FL architecture, the central server can directly observe the clients' reported model updates. 
This may lead to various privacy attacks. 
For example, a colluded server could launch a \emph{membership inference attack} (MIA) \cite{mia_attack} in order to find whether a specific data point was included in \emph{any} client's training dataset.

In this paper, we focus on \emph{source inference attacks} (SIAs) \cite{sia_paper}, which 
aim to identify \emph{exactly which} client owns a data point, in a setting where the central server is honest-but-curious. If successful, a SIA can lead to a severe violation of privacy; for instance, consider a scenario where several hospitals jointly build a medical model using patients' data to treat a disease. 
If an adversary identifies the hospital that owns a particular patient’s record, and that hospital mostly treats COVID-19 patients, the attacker might infer that the patient suffers from COVID-19.

To the best of our knowledge, no effective defense to prevent SIAs has been proposed in the literature. A typical approach in privacy-preserving FL is to use \emph{local differential privacy} (LDP) \cite{ldp_use_1}, where clients perturb their reported model updates by adding noise. However, this approach is not very suitable against a SIA, as it has been shown that the amount of noise necessary to prevent this kind of attacks would significantly deteriorate the accuracy of the joint model \cite{sia_paper}.

\paragraph{Contribution}
In this work, our goal is to design a defense against SIAs that maintains high model accuracy. 
To this aim, we propose \emph{Unary-Quant}; a mechanism involving a trusted shuffler which blends the clients' model updates before releasing them to the central server.
The characteristic of this mechanism is that it does not require the addition of noise. Instead, it uses a unary encoding which, combined with shuffling,  significantly reduces the amount of information available to the central server. To counter the high communication cost of unary encoding, Unary-Quant uses gradient quantization. 

We experimentally evaluate the model accuracy of Unary-Quant on the MNIST dataset. The results show that almost no accuracy is lost, i.e.  the model accuracy is close to that of standard FL. 
Furthermore, we conduct experiments on SIAs. The results indicate that our proposed defense can significantly decrease the effectiveness of a SIA, in the sense that the accuracy of a source inference is reduced to nearly the level of a random guess.

\section{Preliminaries}

\paragraph{Federated Learning \cite{fl_mother_paper}}

Federated learning aims to train a global ML model across $N$ clients, each possessing its own local dataset $\mathcal{D}_i$. 
First, each client $i$ updates the global model $W$ using its local data $\mathcal{D}_i$ to generate an updated model $w_i$. Then, the central server aggregates  the local updates from all clients to form the updated global model: 
    $ W \leftarrow \frac{1}{N} \sum_{i=1}^{N} w_i$ (FedAvg).

\paragraph{Quantization}
In FL, to reduce the communication cost, \emph{quantization} can be used to compress the model updates:

\begin{definition}[\cite{comm_efficiency_in_FL}]  
Let $h = (h_1, \ldots, h_{\lambda})$ be the vector representation of a model parameter $p$. Let $h_{\text{max}} = \max_j (h_j)$ and $ h_{\text{min}} = \min_j (h_j)$.  The compressed version (unbiased estimator) of $h$, denoted by $\tilde{h}$, is: $\tilde{h} = h_{\text{max}}$ w.p. $\frac{h_j - h_{\text{min}}}{h_{\text{max}} - h_{\text{min}}}$ and $\tilde{h} = h_{\text{min}}$ w.p. $\frac{h_{\text{max}} - h_j}{h_{\text{max}} - h_{\text{min}}}$.

%Consider the update $\mathbf{H}_t^i$, let $h = (h_1, \ldots, h_{d_1 \times d_2}) = \text{vec}(\mathbf{H}_t^i)$, and let $h_{\text{max}} = \max_j (h_j), h_{\text{min}} = \min_j (h_j)$. The compressed update of $h$, denoted by $\tilde{h}$ is:

\end{definition}

\begin{comment}
\paragraph{Source Inference Attack}
The honest-but-curious central server can launch an SIA to find the owner of a training record:
\begin{definition} [\textbf{Source Inference Attack} \cite{sia_paper}] Given local optimized model $\theta_k$, a training record $z_1$, source inference aims to infer the posterior probability of $z_1$ belonging to the client $k$:
\begin{equation}
    S(\theta_k, z_1) := \mathbb{P}(s_{1k} = 1 \mid \theta_k, z_1).
\end{equation}
\end{definition}
\end{comment}

\paragraph{Trusted Shuffling}
In this work, we assume the presence of a trusted shuffler which has already been studied as a mean to protect privacy (for instance in the shuffle model of Differential Privacy (DP) \cite{googleESA}). Assuming the existence of a trusted shuffler can be considered as a smaller trust assumption compared to assuming that the central server is trusted since shuffling is a primitive operation that can be performed distributively (using MixNets or Multi-Party Computation) or using trusted hardware \cite{dp_shuffle}.

\section{Protection against the SIA}
First, let us clarify why just using standard (one-message) shuffling is \emph{not} enough to efficiently protect against SIAs.
While shuffling does initially break the link between the client and the model update, in FL the adversary may be able to re-identify each client. 
That is because the adversary might have some statistics over the clients' training datasets, which is often assumed in the literature of FL \cite{mia_attack}.
Hence he can use these statistics to remap the data owner and the reported model update, canceling the effect of the shuffler. 

To overcome this obstacle and effectively blend all model updates, a more sophisticated approach to shuffling is necessary.

\begin{comment}
    While shuffling does initially break the link between the client and the model, in FL the adversary may be able to re-identify each client by using a SIA \Af{what do you mean "by using a SIA"?}. For example, we suppose there are clients A and B are participating in federated learning. Through shuffling, the models' updates are anonymized as $mu_1$ and $mu_2$ \Af{What notation should we stick to for model updates? here $mu_1$ is used} to and the server receives these updates. By performing a SIA using data sampled from the distribution of client A (the adversary may have some statistics over the training datasets of the clients, which is often assumed in the literature of FL \cite{mia_attack}), the server can identify which model update, among $mu_1$ and $mu_2$, belongs to client A. Hence the server can remap the client and the model update, canceling the effect of the shuffler. 
\Af{I dont understand this paragraph, maybe we should simpify. I think the previous way that this paragraph was written was better (I have it as a comment in the Latex). }
\end{comment}

\subsection{A first approach using Unary Encoding}  \label{def1}
To begin with, let us set  aside the communication cost and discuss a simplified variant of Unary-Quant.

The core idea is, informally, that releasing a shuffled bit vector is privacy-wise equivalent to releasing its sum \cite{dp_shuffle}. For example, take a bit vector of length 4 with 2 ones and 2 zeros. The statements: \emph{"the sum of the vector is 2"} and \emph{"the values of the vector (after shuffling) are $\{1,0,1,0\}$"}, provide the adversary with the same amount of information. Observe that this applies only to bit vectors and not, for example, to integer vectors.
However, in reality, most models involve parameters with values in $\mathbb{R}$, which are then typically bounded by clipping. In this work we assume w.l.o.g. that they are clipped in $[-1,1]$ and introduce an encoding step (\Cref{unary_enc_algth}) based on \cite{dp_shuffle}. 

\begin{algorithm}[ht] 
\SetKwInOut{Input}{Input}
\SetKwInOut{Output}{Output}
\SetKwInOut{Return}{Return}

 \Input{$x \in \mathbb{R}$ where $-1 \leq x \leq 1$, $r \in \mathbb{N}$}

\Output{ $(b_1, \ldots, b_r) \in \{0, 1\}^r$}
  \caption{$E(x,r)$: Unary encoding of $x$ \cite{dp_shuffle}}
  \label{unary_enc_algth}
\If{$x = 0$}
       { \textbf{Return} $\{0\}^r$}
    
     $x' \gets (1 + x) / 2$\;
Let $\mu \leftarrow \lceil x' \cdot r \rceil$ and $q \leftarrow x' \cdot r - \mu + 1$

\For {$j = 1, \ldots, r$}{

$b_j = \begin{cases} 
1 & \text{if } j < \mu \\
\operatorname{Ber}(q) & \text{if } j = \mu \\
0 & \text{if } j > \mu 
\end{cases}$
}
\textbf{Return} $(b_1, \ldots, b_r)$
\end{algorithm}
Now consider a mechanism as follows: every client trains their model and encodes every parameter $p$ of the model update to a bit vector $b$ of size $r$ using $E(p,r)$. 
Then, every $b$ is sent to the shuffler. Note that each message should also include some metadata describing what $b$ represents (for example its layer number, if CNN is used).  After all these bit vectors are shuffled, they are released to the central server which can aggregate them and form the joint model. 

Observe that the released output of the shuffler completely prevents the adversary from distinguishing each local model and therefore performing a SIA. 
This is because only a shuffled vector of  bits is available to the adversary. The only information from this vector that is useful to her is its sum, which only allows her to construct the joint aggregated model.

%In contrast, in other works of FL that use the shuffle model, the adversary \emph{does} observe the clients' models under two limitations: a) each model contains some noise and b) he does not know which model belong to which client (because of the shuffling). Nonetheless, as explained before, the power to re-identify or not a client lies in the adversary prior knowledge and in the heterogeneity of the training datasets. 

The Achilles' heel of this approach is its communication complexity. For example, if a CNN is used with $n$ layers and each layer $i$ has $\lambda_i$ parameters, then each client has to send $r \cdot \sum_{i=1}^n \lambda_i$ bits.  Despite the fact that this solution may still be applicable to the so-called \emph{cross-silo} setting of FL, where each client typically has increased communication capabilities, we are about to explore in the following section a variant that decreases the cost while still offering sufficient protection.

\subsection{Unary-Quant} \label{def2}
Quantization can efficiently compress a model update, and since the result is an unbiased estimator of the initial value the impact on the model's accuracy is expected to be negligible.

The core idea of \emph{Unary-Quant} is to use the expensive  approach of \Cref{def1} to transmit only the first $k$ decimal places of each parameter of the model update; the rest can be transmitted through the  cheaper (in terms of communication cost) quantization. In other words, we decompose each parameter $p$ into two segments: $p^a$ and $p^b$ s.t.  $p^a$ contains the first $k$ decimal places of the value and $p^b$ contains the rest. Then unary encoding is used in the part $p^a$ and quantization is used in the part $p^b$. The central server can combine the two parts, after they are shuffled, to form the joint model.  \Cref{alg:quant-shuffle} provides an outline of Unary-Quant and \Cref{alg:fl} shows how it is used in FL.

In essence, the adversary can only use the $p^b$ segment to perform a SIA. Moreover, re-identifying each client only by her $p^b$ is challenging and requires arguably strong assumptions (for example the adversary knowing the clients' corresponding $h_{min}$ and $h_{max}$). Note that in \Cref{alg:quant-shuffle}, we applied $1$-bit quantization, but it can be extended to $n$-bit quantization by dividing   the range $h_{min}$ and $h_{max}$ into $2^n$ equal intervals \cite{comm_efficiency_in_FL}.

%x_1 =int(x)+\frac{\lfloor 10^n frac(x) \rfloor}{10^n}$ and $x_2 = x - x_1$, where  $int(x)$ returns the whole number part and $frac(x)$ returns the fractional part.

\begin{algorithm}
\SetKwInOut{Input}{Input}
\SetKwInOut{Output}{Output}
\SetKwInOut{Return}{Return}

 \Input{$x_j \in w_{t+1}^j$, $r \in \mathbb{N} ,k \in \mathbb{N} $, where $x_j$ has $\lambda$ parameters and each parameter $p$ is $-1 \leq p \leq 1$}

\Output{ $U=(u_1,\ldots,u_{\lambda})$, $H = (h_1, \ldots, h_{\lambda})$ }
\caption{Unary-Quant} 
\label{alg:quant-shuffle}
$h_{max} :=-1$; $h_{min}:=1$\\
\For{each parameter $p_i = p_1 \ldots p_\lambda$ of $x_j$}  
{
    \tcp{Split $p_i$ in parts} 
    
    $p_i^a :=int(p_i)+\frac{\lfloor 10^k frac(p_i) \rfloor}{10^k}$ 
    
    $p_i^b := p_i - p_i^a$ 
    
    \tcp{Unary encoding of $p^a_i$} 
    $U_i \leftarrow E(p^a_i,r)$ 
    %$p_j$=$x_j - (x_j \cdot 10^r)/10^r$\\

    \tcp{Calculate $h_{max}$ and $h_{min}$}
    \If{$p^b_i>h_{max}$}
    {$h_\text{max}$=$p^b_i$}
    \If{$p^b_i<h_{min}$}
    {$h_\text{min}$=$p^b_i$}
}
\tcp{Quantization}
\For{each $p_i^b = p^b_1, \ldots, p^b_\lambda$}  
{

$H_i \leftarrow$ Quantization($p^b_i$,$h_\text{max}$,$h_\text{min}$)
}
\textbf{Return} $U$, $H$

\end{algorithm}

\begin{algorithm}
\SetKwInOut{Input}{Input}
\SetKwInOut{Output}{Output}
\SetKwInOut{Return}{Return}

\Input{Number of rounds $T$, number of clients $N$}
\Output{Final global model $w_R$}
\caption{Federated Learning} 
\label{alg:fl}
\textbf{Initialize} global model $w_0$

\For{each round $t = 1, 2, \ldots, T$}  
{
    \tcp{Server-side}
    Randomly select a subset of clients $S_t$ of size $n \leq N$ \\
    BroadcastGlobalModel($S_t$,$W_t$)\\
    \tcp{Client-side}   
    \For{$j \in S_t$ in parallel} {
        $w_{t+1}^j  \leftarrow W_t - \eta \nabla \ell(W_t; \mathcal{D}_k)$\\   
        $U^j, H^j$ = Unary-Quant($w_{t+1}^j, r,k$)\\
        \text{Send $U^j, H^j$ to shuffler}
    }
    
    \tcp{Shuffler-side}
    \text{Concatenate all $U^j$ and $H^j$ to a single vector $U$ and $H$} \\
    \text{Send Shuffle$(U)$ and Shuffle($H$) to the server}\\
    \tcp{Server-side}
    $W_{t+1} \leftarrow \textbf{FedAvg}(U)+\textbf{FedAvg}(H)$
}

\Return{Final global model $w_R$}

\end{algorithm}

%\Af{What is the commnuciation cost of Unary-Quant? Kangsoo: it's $10^r$ times larger than original shuffling}

%The key difference between Unary-Quant and existing FL approaches lies in the post-training processing of local model parameters on the client side. After training, local model parameters are divided into two components: the encoding part and the quantization part, and then unary encoding and quantization is applied(Algorithm 2). The unary Encoding part produces results that is equivalent to releasing it sum through shuffling, and it is combined with the quantization part to form a joint model on the server side. \Af{I think this paragraph is not necessary since I have already described the algorithm before}

\section {Preliminary Evaluation} \label{evaluation}

In this section, we conduct a preliminary experiment to measure the effectiveness of Unary-Quant, in terms of both model accuracy and preventing SIAs, comparing it to the baseline of standard FL (i.e. without any defense mechanism). We use the MNIST dataset with 10 clients and use a Dirichlet distribution (setting its hyperparameter $\alpha$ to $0.1$) to simulate the heterogeneity of the training data.  We use a CNN model and the total number of model parameters is 421642.

\begin{figure}[h]
		\centering
		\includegraphics[width=0.85 \columnwidth]{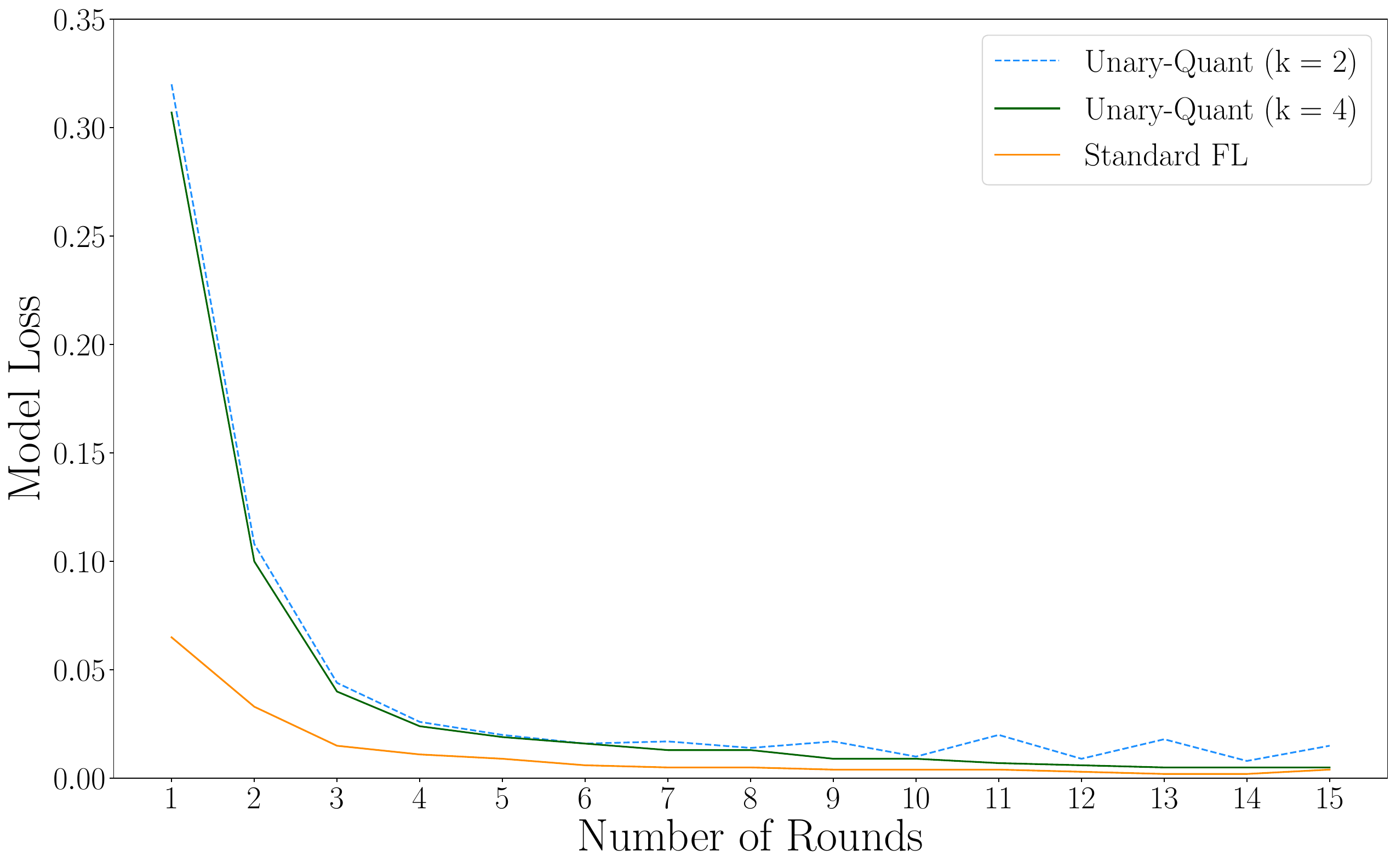}
		\caption{Model Loss}
		\label{fig:model_loss}
    \Description{Standard FL model loss is around 0.01 when the number of rounds is 15, slightly less than Unary-Quant for both k=2 and k=4}
\end{figure}

First we measure the model loss using Unary-Quant with $k=2$ and $k=4$ while setting $r = 10^k$; \Cref{fig:model_loss} shows that in both cases the model loss quickly approaches that of standard FL  as the number of rounds increase. 
\Cref{table:model_accuracy} shows that Unary-Quant achieves model accuracy nearly identical to standard FL while effectively protecting against SIAs: reducing their accuracy from $44.5\%$ to $14.7 \%$.
Recall that the baseline of random guess
is $10\%$ (assumed to be uniform over all clients). 

\begin{table}[h] 
  \centering
  \begin{tabular}{@{}ccc@{}}
    \toprule
    Method  & Model Accuracy & SIA accuracy \\ 
    \midrule
    Standard FL  & 98.8  & 44.5 \\ 
    \hline
    Unary-Quant ($k=3$, $r=10^3$)  & 98.1 & 14.7 \\  
    \bottomrule
  \end{tabular}
  \caption{Model and SIA accuracy after 15 rounds (percentage)}
    \label{table:model_accuracy}
\end{table}

\section{Discussion}

The benefit of our approach is that it is primarily based on encoding, allowing for direct integration with other methods in FL that already use a trusted shuffler (e.g. the shuffle model of DP \cite{shuffle_dp_FL}). 
Our experiments indicate that Unary-Quant achieves model accuracy similar to that of standard FL while notably protecting against SIAs.
More experiments should follow, measuring its effectiveness across multiple datasets with varying parameters (e.g. degree of heterogeneity, number of clients). 
Finally it is vital to explore additional gradient compression techniques as to further reduce the communication cost. 

\begin{acks}
   The work of Andreas Athanasiou was supported by the project CRYPTECS, funded by the ANR (project number ANR-20-CYAL-0006)	  and by the BMBF (project number 16KIS1439).
The work of Kangsoo Jung was supported by the project ELSA, funded by the Horizon Europe Framework (project number 101070617). The work of Catuscia Palamidessi was supported by the project HYPATIA, funded by the ERC (grant agreement number 835294).

\end{acks}

\bibliographystyle{ACM-Reference-Format}
\bibliography{bibliography}

\appendix

\end{document}